\documentclass[12pt]{article}
\usepackage{amsfonts,latexsym}
\begin{document}
\newcommand{\de}{\delta}\newcommand{\ga}{\gamma}
\newcommand{\e}{\epsilon} \newcommand{\ot}{\otimes}
\newcommand{\be}{\begin{equation}} \newcommand{\ee}{\end{equation}}
\newcommand{\ba}{\begin{eqnarray}} \newcommand{\ea}{\end{eqnarray}}
\newcommand{\tmod}{{\cal T}}\newcommand{\amod}{{\cal A}}
\newcommand{\bemod}{{\cal B}}\newcommand{\cmod}{{\cal C}}
\newcommand{\dmod}{{\cal D}}\newcommand{\hmod}{{\cal H}}
\newcommand{\s}{\scriptstyle}\newcommand{\tr}{{\rm tr}}
\newcommand{\einsop}{{\bf 1}}
\def\oR{R^*} \def\upa{\uparrow}
\def\R{\overline{R}} \def\doa{\downarrow}
\def\oL{\overline{\Lambda}}
\def\nn{\nonumber} \def\dag{\dagger}
\def\beq{\begin{equation}}
\def\eeq{\end{equation}}
\def\bea{\begin{eqnarray}}
\def\eea{\end{eqnarray}}
\def\ve{\epsilon}
\def\si{\sigma}
\def\th{\theta}
\def\d{\delta}
\def\ga{\gamma}
\def\l{\left}
\def\r{\right}
\def\a{\alpha}
\def\b{\beta}
\def\g{\gamma}
\def\La{\Lambda}
\def\w{\overline{w}}
\def\u{\overline{u}}
\def\o{\overline}
\def\rr{\mathcal{R}}
\def\T{\mathcal{T}}
\def\N{\overline{N}}
\def\Q{\overline{Q}}
\def\L{\mathcal{L}}
\def\m{\overline{m}}
\def\n{\overline{n}}
\def\p{\overline{p}}
\def\l{\overline{l}}
\def\d{\dagger}
\newcommand{\reff}[1]{eq.~(\ref{#1})} 

\vspace{6cm}

\begin{center}
{\Large \bf Integrabilty and exact spectrum of a pairing model for
nucleons
}
\vskip.3in
{\large Jon Links, Huan-Qiang Zhou, Mark D. Gould and Ross H. McKenzie}
\vskip.2in
{\em Centre for Mathematical Physics, \\
The University of Queensland,
      4072, \\ Australia

      Email: jrl@maths.uq.edu.au}
      \end{center}

      \vskip 2cm
      \begin{abstract}

A pairing model for nucleons, introduced by Richardson in 1966, which
describes proton-neutron pairing as well as proton-proton and
neutron-neutron
pairing, is re-examined in the context of the Quantum Inverse Scattering
Method. Specifically, this shows that the model is integrable by enabling
the explicit 
construction of the conserved operators. We determine the eigenvalues
of these operators in terms of the Bethe ansatz, which in turn leads to
an expression for the energy eigenvalues of the Hamiltonian.   
      \end{abstract}          

\vfil\eject

\vspace{1cm}
    \centerline{\bf 1. Introduction.}
\vspace{1cm}     
Pairing model Hamiltonians have again become the focus of many theoretical
condensed matter investigations  due to the fact
that experimental work on metallic nanoparticles (~also refered to as
small metallic grains~) has detected evidence of
pairing interactions \cite{rbt}. In order to gain an insight into the
physical properties of small metallic grains, substantial attention has
been devoted to the analysis of the (~reduced~) BCS Hamiltonian which is
believed to be
appropriate to describe the dynamics of these systems \cite{dr01}. An
important point in this regard is
that the treatment originally proposed by Bardeen, Cooper and Schrieffer
for bulk systems,
using variational wavefunctions with an undetermined number of particles
(~grand canonical ensemble~), is not applicable to the study of a small
metallic grain where the number
of electrons remains fixed (~canonical ensemble~). This aspect has
generated  
activity in
analysing the BCS Hamiltonian under this constraint \cite{bv}. Remarkably,
an exact
solution of the reduced BCS Hamitonian was obtained some time ago in a
series of
works by Richardson and Sherman \cite{r63}. In these papers, the BCS
Hamiltonian was studied for the purpose of application to
pairing interactions in nuclear systems, and as such the solution
escaped the attention of the condensed matter physics community for a
considerable time. More recently, it was
shown by Cambiaggio, Rivas and Saraceno \cite{crs97} that the model is 
also
integrable; i.e. there exists a set of mutually commuting operators which
commute with the Hamitonian. These features can be reproduced in
the framework of the Quantum Inverse Scattering Method (~QISM~) using a
solution of the Yang-Baxter equation as shown in \cite{zlmg,vp}. This
approach
has the significant 
advantage that the computation of form factors and correlation
functions can be undertaken in this algebraic framework
\cite{zlmg,ao01}. 

In the papers \cite{r,r1} Richardson introduced a coupled pairing model 
for nuclear systems which
accomodates proton-neutron pairing interactions as well as
the proton-proton
and neutron-neutron couplings. This model was recently studied in
\cite{dmnss}. In this paper we will show that the
techniques employed in \cite{zlmg,vp} can be applied to this model to
show
that it is integrable, and for the determination of the energy spectrum.
This formulation also opens the possibility for the calculation 
of form factors and correlation functions by algebraic means, in analogy 
with the results of \cite{zlmg}. 
For the present case 
the solution is obtained
through the use of a solution of the Yang-Baxter equation
associated with the Lie algebra $so(5)$.  
The Hamiltonian has the explicit form 
\beq H=\sum_j^D\e_jn_j
-g\sum_{j,k}^D\left(b_j^{\d}(1)b_k(1)+b_j^{\d}(2)b_k(2)+b_j^{\d}(3)b_k(3)
\right)
\nn
\label{ham} \eeq 
where $g$ is an arbitrary coupling parameter and $D$
is the total number of distinct energy levels. Above,
$n_j$ is the number operator for
paired nucleons at energy level $\e_j$ and
$b(i)_j,\,b^{\d}_j(i),\,\,i=1,2,3$
are the annihilation and creation
operators for three sets of generalized, non-commuting, hard core boson
operators
satisfying the relations (amongst others) 
$$\left(b^{\d}_j(i)\right)^2=0, ~~i=1,2,~~~~~~
\left(b^{\d}_j(3)\right)^3=0
$$
$$[b_j(i),\,b_k(l)]=[b^{\d}_j(i),\,b^{\d}_k(l)]=
[b_j(i),\,b^{\d}_k(l)]=0, ~~~{\rm for }~~ k\neq
j. $$    
The three sets of hard core boson operators correspond to proton-proton,
neutron-neutron and proton-neutron pairing in a nuclear system. Their
explicit forms will now be made clear. Two sets of two-fold
degenerate Fermi operators $c_{\pm}$ and $d_{\pm}$ and their
hermitian conjugates are introduced.
$c_{\pm}$ and $d_{\pm}$ represent the protons and neutrons, 
the subscripts $\pm$ referring to time reversed
states. The hard-core bosons are realized by 
\bea
b^{\dagger}(1)&=&c^{\dagger}_+c^{\dagger}_-, \nn \\
b^{\dagger}(2)&=&d^{\dagger}_+d^{\dagger}_-, \nn \\
b^{\d}(3)&=&\frac{1}{\sqrt 2}\left(c^{\d}_+d^{\d}_-+d^{\d}_+c^{\d}_-\right)
\nn \eea  
with appropriate definitions for the hermitian conjugates.
The Hamiltonian (\ref{ham}) corresponds to a special case where the
energy level spacings for the protons and neutrons are the same and the
scattering coupling is the same for proton-proton, neutron-neutron and
proton-neutron pairings. In this instance the Hamiltonian acquires an
{\it isospin} symmetry which plays an important role in our analysis
below.

The energy
levels $\e_j$ are degenerate. Each level can be empty or occupied by
protons and/or neutrons in two-fold degenerate time reversed states.  
This gives the degeneracy of each level as $2^4=16$. However,
the above Hamiltonian only scatters paired nucleons giving
rise to a blocking effect (~c.f., the blocking effect for the BCS model
discussed in \cite{dr01}~). For any energy level which contains an odd
number of nucleons, the pairing interaction acts trivially, and these
states can be discarded from the Hilbert space. (~Hence, we
choose $n_j$ to count the number of {\it paired} nucleons at
$\e_j$, rather than the number of nucleons.~) Furthermore, it is assumed
that the proton-neutron
pairing is between time reversed states which are symmetric        
under interchange of protons and neutrons. In this case the pairing
interaction 
is also trivial on the non-time reversed paired proton-neutron
states and the antisymmetric time reversed paired proton-neutron state,
giving
rise to a 5-dimensional space at each level $\e_j$ on which the
scattering is non-trivial. (~In the language of \cite{r}, this Hilbert space 
is spanned by the {\it Seniority-Zero} states. In the subsequent paper 
\cite{r1} Richardson extended his results to Seniority-One and -Two states. 
However, we will not consider these cases here.~) 
It is convenient to use the fundamental
representation of the $so(5)$ Lie algebra to construct the local operators
acting on each of these spaces,  which we will now discuss. 

\vspace{1cm} 
    \centerline{\bf 2. The Lie algebra $so(5)$.}
\vspace{1cm}   
We can construct the fundamental representation of the $so(5)$
Lie algebra in the following manner. Define $\m=6-m$. For $5\times 5$
matrices, consider 
the subset 
$$a^m_n=e^m_{\n}-e^n_{\m}=-a^n_m$$
where $e^m_n$ denotes the matrix with 1 in the $(m,n)$ position and
zeroes elsewhere. Note that $a^m_m=0$. We will denote the 5-dimensional
space on which
these operators act by $V$. 
The operators $a^m_n$ close to form the fundamental or
defining representation of the Lie algebra
$so(5)$ 
with commutation relations
\beq [a^m_n,\,a^p_l]=\de^p_{\n}a^m_l+\de^l_{\m}a^n_p
+\de^n_{\l}a^p_m+\de^m_{\p}a^l_n \label{cr}. \eeq
A basis for the Lie algebra is given by the set
$$\{a^m_n:1\leq m<n\leq 5\}$$  
which gives ten linearly independent generators. Explicitly the basis
generators read 
\bea &&a^1_2=e^1_4-e^2_5, 
~~~a^1_3=e^1_3-e^3_5, \nn \\
&&a^1_4=e^1_2-e^4_5, 
~~~a^1_5=e^1_1-e^5_5, \nn \\
&&a^2_3=e^2_3-e^3_4, 
~~~a^2_4=e^2_2-e^4_4, \nn \\
&&a^2_5=e^2_1-e^5_4, 
~~~a^3_4=e^3_2-e^4_3, \nn \\
&&a^3_5=e^3_1-e^5_3, 
~~~a^4_5=e^4_1-e^5_2. \nn \eea 
Note that the representation is 
unitary, and specifically 
\beq \left(a^m_n\right)^{\dagger}=a^{\n}_{\m}.
\label{hc} \eeq 

Next, we recall some established results on the representation theory of
$so(5)$. For a more detailed discussion, see for example \cite{br}.
Finite dimensional irreducible representations of $so(5)$ are uniquely
determined by the highest weight labels $\Lambda_1,\,\Lambda_2$ which are
the eigenvalues of the Cartan elements 
$$h_1=a^1_5,~~~~h_2=a^2_4$$ 
acting on the highest weight state. (~These operators are self-adjoint and
mutually commuting and so can be diagonalized simultaneously.~) The
highest weight state is the unique
vector $v$ which vanishes under the action of the raising operators; viz, 
$$a^1_2v=a^1_3v=a^1_4v=a^2_3v=0. $$ 
The highest weight labels $\Lambda_1,\,\Lambda_2$ take integer or
half odd-integer values
and are subject to the
conditions
$$\Lambda_1\geq \Lambda_2\geq 0,
~~~~~~\Lambda_1-\Lambda_2\in\mathbb{Z}.$$ 
In
the case of the fundamental representation we have
$\Lambda_1=1,\,\Lambda_2=0$ and the highest weight vector is
$\left|1\right>\equiv (1,0,0,0,0)^t$, with $t$ the matrix transposition
operation. 

The $so(5)$ algebra admits a second order Casimir invariant
\beq C=\sum_{m,n}^5 a^m_na^{\n}_{\m} \label{cas} \eeq
commuting with all elements of $so(5)$, which can be verified
explicitly from the commutation relations (\ref{cr}).    
Due to Schur´s lemma, on each irreducible finite dimensional
representation the Casimir element
(\ref{cas}) takes a constant  eigenvalue, which is  given by  
$$\chi_C(\Lambda_1,\,\Lambda_2)=2\left(\Lambda_1(\Lambda_1+3) + 
\Lambda_2(\Lambda_2+1)\right). $$ 
For the fundamental representation we clearly have $\chi_C(1,\,0)=8.$ 

An important ingredient in the following construction is the
existence of a canonical 
$so(3)$ subalgebra
spanned by the basis elements 
\bea &&L^0=a^2_4, 
~~~L^+=a^2_3, 
~~~L^-=a^3_4. \nn \eea
The Casimir operator for this $so(3)$ subalgebra is given by 
$$K=L^+L^-+L^-L^++(L^0)^2. $$ 
The irreducible finite dimensional representations of the $so(3)$ algebra
have a unique highest weight  
vector $w$
which satisfies $L^+w=0$. These representations are uniquely
characterized by the eigenvalue $\mu$ of $L^0$ acting on $w$. The
allowable values of $\mu$ are such that
it is a non-negative integer or half odd-integer. The eigenvalue of $K$
on such a
representation is given by 
$$\chi_K(\mu)=\mu(\mu+1). $$  
The realization of this 
canonical $so(3)$ subalgebra we give below 
is referred to as the isospin algebra
in \cite{r}. 

\vspace{1cm}
    \centerline{\bf 3. The Yang-Baxter equation and integrability.}
\vspace{1cm}              

The basis for constructing an integrable model through the QISM \cite{fad}
is a solution of the Yang-Baxter equation 
\beq 
R_{12}(u-v)R_{13}(u)R_{23}(v)=R_{23}(v)R_{13}(u)R_{12}(u-v)
\label{yb} \eeq  
which is a matrix solution acting on a three-fold tensor product
space $V\otimes V \otimes V$. Above, the subscripts refer to which 
two of the three spaces the operator $R(u)\in {\rm End} (V\otimes
V$) acts upon. Solutions of the Yang-Baxter
equation associated with representations of Lie algebras are well known. 
The $R$-matrix solution associated with the fundamental  representation of
$so(5)$ discussed above 
takes the 
following form \cite{cgx}: set 
\bea &&I=\sum^5_{m,n}e^m_m\otimes e^n_n, \nn \\
&&P=\sum^5_{m,n}e^m_n\otimes e^n_m, \nn \\
&&Q=\sum^5_{m,n}e^m_n\otimes e^{\m}_{\n}. \nn \eea 
Then 
$$R(u)=I+\frac{2\eta}{u}P-\frac{2\eta}{u+3\eta}Q $$ 
provides a solution of (\ref{yb}) with $\eta$ being a free parameter. This
solution is $so(5)$ invariant in 
that 
$$[R(u),\,I\otimes x+x \otimes I]=0$$ 
for any $x\in so(5)$. We note the properties:

1. Unitarity

$$R(u)R(-u)=\left(\frac{u^2-4\eta}{u^2}\right)I\otimes I. $$  

2. Crossing symmetry

$$R^{t_1}(-u-3\eta)=(I\otimes A)R(u)(I\otimes A)$$ 
where $A$ is the  matrix with elements $A^m_n=\de^m_{\n}$ and $t_1$ denotes 
matrix transposition in the first space of the tensor product. 

By the 
usual procedure of the QISM we define a transfer matrix acting on the
$D$-fold
tensor product space via
$$t(u)=tr_0\left(G_0R_{0D}(u-\e_D)..........R_{01}(u-\e_1)\right)$$ 
which gives a commuting family satisfying $[t(u),\,t(v)]=0$. Above, 
$tr_0$ denotes the trace taken over an auxiliary space labelled by 0
and $G$ can be any matrix which satisfies 
$$[R(u),\, G\otimes G]=0.$$ 
We choose $G=\exp(\alpha\eta a^1_5)$ for this 
particular model. 
Using
either the analytic Bethe ansatz, which exploits the unitarity and
crossing symmetry properties \cite{resh}, or the algebraic Bethe 
ansatz developed by Martins and Ramos for the $so(n)$ series \cite{mr},
the
eigenvalues of the transfer matrix are found to be 
\bea  
\Lambda(u)&=&\exp(\alpha\eta)\prod_k^D\frac{(u-\e_k+2\eta)}{(u-\e_k)} 
\prod_i^M\frac{(u-v_i-\eta)}{(u-v_i+\eta)} 
\nn \\
&&+\exp(-\alpha\eta)\prod_k^D\frac{(u-\e_k+\eta)}{(u-\e_k+3\eta)}
\prod_i^M\frac{(u-v_i+4\eta)}{(u-v_i+2\eta)}+\Lambda_0(u) \nn \eea  
where $\Lambda_0(u)$ are the transfer matrix eigenvalues for the $R$-matrix 
associated with the spin 1 Babujian-Tahktajan model \cite{bt}, with
inhomogeneities 
$v_i$. These eigenvalues read 
\bea 
\Lambda_0(u)&=&\prod_i^M\frac{(u-v_i+3\eta)}{(u-v_i+\eta)}
\prod_j^N\frac{(u-w_j)}{(u-w_j+2\eta)} 
\nn \\
&&~+\prod_i^M\frac{(u-v_i)}{(u-v_i+2\eta)}
\prod_j^N\frac{(u-w_j+3\eta)}{(u-w_j+\eta)} \nn \\
&&~~+\prod_j^N\frac{(u-w_j+3\eta)}{(u-w_j+\eta)}\frac{(u-w_j)}{(u-w_j+2\eta)}. 
\nn \eea 
The parameters $v_i, w_j$ are required to satisfy the Bethe ansatz equations
\bea 
\exp(\alpha\eta)\prod_k^D\frac{(v_j-\e_k+\eta)}{(v_j-\e_k-\eta)}
&=&-\prod_l^N\frac{(v_j-w_l-\eta)}{(v_j-w_l+\eta)}
\prod_i^M\frac{(v_j-v_i+2\eta)}{(v_j-v_i-2\eta)}, \nn \\
\prod_i^M\frac{(w_j-v_i-\eta)}{(w_j-v_i+\eta)}
&=&-\prod_k^N\frac{(w_j-w_k-\eta)}{(w_j-w_k+\eta)}. \nn \eea 
 
Define the operators 
\beq T_j=\lim_{u\rightarrow\e_j}\frac{u-\e_j}{2\eta}t(u)\label{T}\eeq 
which satisfy 
\beq [T_j,\,T_k]=0. \label{com1} \eeq  
Explicitly
$$T_j=G_iR_{jD}(\e_j-\e_D)...R_{j(j+1)}(\e_j-\e_{j+1})R_{j(j-1)}(\e_j-\e_{j-1})
...R_{j1}(\e_j-\e_1). $$ 
Now, by  taking the {\it quasi-classical} expansion 
\beq T_j=I+\eta\tau_j+o(\eta^2)\label{Texp}\eeq  
we find 
$$\tau_j=\alpha\psi_j+2\sum^D_{k\neq j}\frac{\phi_{jk}}{\e_j-\e_k}$$
where 
$$\phi=\sum^5_{m,n} e^m_n\otimes a^n_{\m}$$ 
and for ease of notation we set $\psi=a^1_5$. Setting
\bea 
\theta&=&\sum^5_{m,n} a^m_n\otimes a^{\n}_{\m} \nn \\
&=&\sum^5_{m,n}(e^m_{\n}-e^n_{\m})\otimes a^{\n}_{\m} \nn \\
&=&\sum^5_{m,n}e^m_n\otimes a^n_{\m} +\sum_{m,n}e^n_{\m}\otimes
a^{\m}_{\n}\nn\\
&=&2\sum^5_{m,n}e^m_n\otimes a^n_{\m}\nn \\
&=&2\phi \nn  \eea 
shows that we  may  write 
$$\tau_j=\alpha\psi_j+\sum^D_{k\neq j}\frac{\theta_{jk}}{\e_j-\e_k}$$
which satisfy 
$[\tau_j,\,\tau_k]=0 $ 
in view of (\ref{com1},\ref{Texp}).

Consider the action of the $so(3)$ Casimir on the $D$-fold tensor product
space
$$K\rightarrow \sum^D_{i,j}(L^+_iL^-_j+L^-_iL^+_j+L^0_iL^0_j).$$
It is easily deduced that 
$$[K,\, \psi_j]=0.$$ 
When $\alpha=0$ ( and so $G=I$ ),  
$$[K,\, \tau_j]=0$$ 
since in this instance the operators $\tau_j$ are $so(5)$ invariant as a 
consequence of the $so(5)$ invariance of the $R$-matrix. We thus see that 
$$[K,\,\tau_j]=0$$ 
in general. 
The set of operators
$\{\tau_i,\,K\}$ are mutually commutative
and so can be used to define an integrable Hamiltonian through any 
function
of these operators. With an appropriate choice, we show below that the
pairing Hamiltonian (\ref{ham}) introduced in \cite{r} can be reproduced,
thus
establishing
the integrability of this model.  

\vspace{1cm}
    \centerline{\bf 4. Pairing Hamiltonian.}
\vspace{1cm}              
First, let us realize the 5-dimensional space $V$ in terms of the two sets of 
two-fold degenerate Fermi 
operators $c_{\pm}, \,c^{\d}_{\pm}$ and $d_{\pm},\,d^{\d}_{\pm}$ 
introduced earlier.   
We make the identifications
\bea 
\left|1\right>
&=&\left|0\right>, \nn \\
\left|2\right>
&=&d^{\dagger}_+d^{\dagger}_-\left|0\right>, \nn \\
\left|3\right>
&=&\frac{1}{\sqrt 2}\left(c^{\dagger}_+d^{\dagger}_-
+d^{\dagger}_+c^{\dagger}_-\right)\left|0\right>, \nn \\
\left|4\right>
&=&c^{\dagger}_-c^{\dagger}_+\left|0\right>, \nn \\
\left|5\right>
&=&
c^{\dagger}_+c^{\dagger}_-d^{\dagger}_+d^{\dagger}_-
\left|0\right>. \nn \eea 
Set $n^c=c^{\dagger}_+c_++c^{\dagger}_-c_-,\,n^d=d^{\dagger}_+d_+
+d^{\dagger}_-d_-$ and $n=1/2(n^c+n^d)$,  which measures the number of
paired fermions. 
We have the following realisation of the $so(5)$
generators 
\bea 
a^4_5&=&c^{\dagger}_-c^{\dagger}_+, \nn \\
a^2_5&=&d^{\dagger}_+d^{\dagger}_-, \nn \\
a^3_4&=&\frac{1}{\sqrt 2}\left(c^{\dagger}_-d_-+c^{\d}_+d_+\right), \nn \\
a^3_5&=&\frac{1}{\sqrt 2}\left(c^{\d}_+d^{\d}_-+d^{\d}_+c^{\d}_-\right), \nn \\
a^1_5&=&I-\frac12\left(n^c+n^d\right), \nn \\
a^2_4&=&\frac12\left(n^d-n^c\right) \nn \eea 
and 
$$a^1_4=(a^2_5)^{\d},~~a^1_3=(a^3_5)^{\d},~~a^2_3=(a^3_4)^{\d}, 
~~a^1_2=(a^4_5)^{\d}. $$ 

The representation of the canonical $so(3)$ subalgebra generated by
$\{L^+,\,L^-,\,L^0\}$ is the isospin algebra of \cite{r}, and the
operator $\psi$ is a $U(1)$ generator.
We can also identify the   
generalized hard core boson operators with certain elements of the $so(5)$
algebra
through
$$ 
b^{\d}(1)=-a^4_5,~~~b^{\d}(2)=a^2_5,~~~b^{\d}(3)=a^3_5 $$ 
and corresponding relations for the hermitian conjugates from (\ref{hc}). 
We may now express
\bea 
\frac12\theta&=&b^{\d}(1)\otimes b(1)+ 
+b^{\d}(2)\otimes b(2)
+b^{\d}(3)\otimes b(3) \nn \\
&&+b(1)\otimes b^{\d}(1)
+b(2)\otimes b^{\d}(2)
+b(3)\otimes b^{\d}(3) \nn \\
&&+L^+\otimes L^-+L^-\otimes L^++ L^0\otimes L^0 +\psi\otimes\psi. \nn \eea 

Define the Hamiltonian 
\bea 
H&=&\frac{-1}{\alpha}\left(\sum_j^D(\e_j-\frac{3}{\alpha})\tau_j-K+4D\right)
+\frac{1}{\alpha^3}\sum_{j,k}^D\tau_j\tau_k
 +\sum_j^D\e_j\nn \\
&=&-\sum_j^D(\e_j-\frac{3}{\alpha})\psi_j
-\frac{1}{2\alpha}\sum_j^D\sum^D_{k\neq j}\theta_{jk}
+\frac{1}{\alpha}\left(\sum_{j,k}^D\psi_j\psi_k + K-4D\right)
+\sum_j^D\e_j\nn 
\\
&=&-\sum_j^D(\e_j-\frac{3}{\alpha})\psi_j
-\frac{1}{2\alpha}\sum^D_{j,k}\theta_{jk}
+\frac{1}{\alpha}\left(\sum_{j,k}^D\psi_j\psi_k + K\right)
+\sum_j^D\e_j\nn \eea 
where in the last line we have use the fact that the so(5) Casimir invariant
$C$
takes the eigenvalue 8 in the 5-dimensional representation, as mentioned
earlier. 
Expressing  the $so(5)$ elements in terms of the hard core boson
operators
as
indicated above yields (\ref{ham}) with $g=2/\alpha$.   
The Hamiltonian 
describes two coupled identical 
BCS models,  where, in addition to the customary pairing  
(~characterized by the operators 
$b^{\d}(1),\, b(1),\,b^{\d}(2),\, b(2)$~),  
fermions from each BCS system 
at the same energy level $\e_j$ and in time reversed states can pair in 
such a fashion that the wave function is symmetric under interchange of the 
two BCS systems (~described by $b^{\d}(3),\,b(3)$~). This shows that the
Hamiltonian has a natural
interpretation as a
pairing model for nucleons which includes proton-neutron pairing.  

\vspace{1cm}
    \centerline{\bf 5. Energy Spectrum}
\vspace{1cm} 

It remains to determine the eigenvalues of the Hamiltonian, which is 
achieved by computing the eigenvalues of the conserved operators. 
The eigenstates can be labelled by the eigenvalues of the Cartan elements
$h_1=a^1_5=\Psi,\,h_2=a^2_4=L^0$ 
acting on the tensor product space. For each value 
of $M$ and $N$ which appear in the  solution of the Bethe ansatz equations, 
we find that the corresponding eigenvalues of the Cartan elements are 
$$ \Lambda_1=D-M, ~~~~\Lambda_2=M-N=\mu $$ 
or equivalently
$$n=M,~~~~n^d-n^c=2(M-N). $$ 
We begin with the operator $K$. It follows from the algebraic
construction of the Bethe states due to Martins and Ramos \cite{mr} 
that each Bethe state is a highest weight state
with respect to the $so(3)$ subalgebra.   
(~When $\alpha=0$ the Bethe states are 
highest weight states with respect to the full $so(5)$ algebra, but
generic values of $\alpha$ break this symmetry.~) Since $K$ is 
simply the $so(3)$ Casimir operator, it takes the eigenvalue 
$\chi_K(M-N)=(M-N)(M-N+1)$ 
on such a Bethe state.

From (\ref{T},\ref{Texp}) we see that that the eigenvalues for $\tau_j$
can be obtained from the quasi-classical limit of the transfer matrix
eigenvalues. These read
\beq \lambda_j=\alpha+\sum_{k\neq j}^D\frac{2}{\e_j-\e_k}
-\sum_i^M\frac{2}{\e_j-v_i} \label{eig} \eeq 
and the Bethe ansatz equations take the form
\bea 
\alpha+\sum_k^D\frac{2}{v_j-\e_k}&=&\sum^M_{i\neq j}\frac{4}{v_j-v_i}
+\sum_l^N\frac{2}{w_l-v_j}, \nn \\
\sum_i^M\frac{1}{w_j-v_i}&=&\sum^N_{k\neq j}\frac{1}{w_j-w_k}. \label{bae} 
\eea 
Using these Bethe ansatz equations we can derive the following identities
\bea 
\sum_j^N\sum_i^M\frac{1}{w_j-v_i}&=&\sum_j^N\sum^N_{k\neq
j}\frac{1}{w_j-w_k}
\nn \\
&=&0, 
\nn \\
\alpha M+\sum_j^M\sum_k^D\frac{2}{v_j-\e_k}
&=&\sum_j^M\sum_{i\neq j}^M\frac{4}{v_i-v_j}
+\sum_j^M\sum_l^N\frac{2}{w_l-v_j} \nn \\
&=& 0, \nn \\ 
\sum_j^M\sum_l^N\frac{w_l}{w_l-v_j}-\sum_j^M\sum_l^N\frac{v_j}{w_l-v_j}&=&MN, 
\nn \\
\sum_j^N\sum_i^M\frac{w_j}{w_j-v_i}
&=&\sum_j^N\sum_{k\neq j}^N\frac{w_j}{w_j-w_k} \nn \\
&=&\frac12 N(N-1), \nn \\
\sum_j^M\sum_k^D\frac{v_j}{v_j-\e_k}-\sum_j^M\sum_k^D\frac{\e_k}{v_j-\e_k}
&=&ML, \nn \\
\alpha\sum_j^M v_j+\sum_j^M\sum_k^D\frac{2v_j}{v_j-\e_k}
&=&\sum_j^M\sum^M_{i\neq j}\frac{4v_j}{v_j-v_i}
+\sum_j^M\sum_l^N\frac{2v_j}{w_l-v_j} \nn \\
&=&
2M(M-1)-2MN+N(N-1).\nn 
\eea
We now obtain from these identities 
\bea \sum_j^D\lambda_j&=&\alpha(D-M), \nn \\
\sum_j^D\e_j \lambda_j&=&\alpha\sum_j^D\e_j+D(D-1)
-\alpha\sum_j^Mv_j+2M(M-1)+N(N-1)-2M(D+N). \nn \eea 
Using these results we find that the energies are given by 
\bea 
E&=&\frac{-1}{\alpha}\left(\sum_j^D\e_j\lambda_j-3(D-M)-(M-N)(M-N+1)+4D 
\right)+\frac{1}{\alpha^3}\sum_{j,k}^D\lambda_j\lambda_k
+\sum_j^D\e_j\nn \\
&=& \sum_j^Mv_j.       
\nn \eea 
This energy expression shows that the roots $\{v_j\}$ of the Bethe 
ansatz equations are simply the quasi-particle excitation energies. 
It is interesting to note in the case $N=0\Rightarrow n^c=0$, 
where the model describes a single reduced BCS system since there 
is only one type of nucleon, the 
Bethe ansatz equations and energy expression coincide with those 
obtained by Richardson and Sherman \cite{r63}. 

Finally, it is necessary to compare our results with those obtained 
by Richardson \cite{r}, in which the following Bethe ansatz equations 
were obtained 
\bea
\frac{1}{g}+\sum_k^D\frac{1}{v_j-\e_k}
&=&\frac{M(M-3)+(M-N)(M-N+1)}{M(M-1)}\sum^M_{i\neq j}\frac{1}{v_j-v_i}
\label{bae1}
\eea
and are obviously different from our results. The explanation for this 
difference stems from the fact that the ans\"atze adopted for the 
eigenstate wavefunctions are different in each case. Richardson chose 
wavefunctions which have eigenvalue zero under the action of the isospin
operator $L^0$. As we have indicated above, the states we construct are 
highest weight states with respect to the isospin algebra. An important 
open problem is to prove the equivalence of these two solutions.

\vspace{1cm}
    \centerline{\bf 6. Conclusion.}
\vspace{1cm} 

We have shown, by using the QISM,  that the coupled BCS Hamiltonian
proposed by Richardson
\cite{r} to accomodate proton-neutron pairing in nuclear systems is
integrable. We
have also determined expressions for the energy
eigenvalues of the model in terms of a Bethe ansatz solution of coupled
equations. It should be emphasized that although the model studied here  
is based on a specific Lie algebra and representation, the construction
that we have employed to demonstrate integrability is entirely general. 
It can be equally applied to
any representation of any Lie algebra or superalgebra, to yield a vast
class of integrable systems. For a recent example based on the spin 1
representation of the $so(3)$ algebra see \cite{lzmg}.

An interesting question to ask is whether this solution is
complete; i.e., are all energy levels obtained? It is well known that for
many Bethe ansatz solvable models where there is an underlying Lie algebra
symmetry the eigenstates are highest weight states with respect to this
algebra \cite{ft,k,eks}. By computing the dimensions of each multiplet
generated by these highest weight states and then employing a
combinatorial
argument, completeness can be proved.

For the present model, where the
$R$-matrix
solution
has $so(5)$
symmetry, this symmetry is broken in the construction of the transfer
matrix by the inclusion of the operator $G$, and an $so(3)$ symmetry
(~isospin~) remains for the conserved operators $\tau_j$. 
The degeneracies of the
eigenvalues can be classified in terms of $so(3)$ multiplets. In the
$\alpha\rightarrow 0$ limit
the $so(5)$ symmetry is restored and we find that the Bethe states are
$so(5)$ highest weight states, so there is an increase in the  
degeneracy of each eigenvalue
at this point, or equivalently, a decrease in the number of distinct
eigenvalues. Fortunately, the Bethe ansatz equations (\ref{bae}) 
admit more
solutions for generic values of $\alpha$ than the $\alpha=0$ case and
automatically accomodate this facet. This is
easily illustrated in the instance $D=2,\,M=1,\,N=0$, in which case we
need only solve 
\beq \alpha+\frac{2}{v-\e_1}+\frac{2}{v-\e_2}=0. \label{ex} \eeq  
For non-zero $\alpha$, this is a quadratic equation with two solutions 
for $v$. When $\alpha$ is zero, the equation is linear with the
unique finite solution $v=(\e_1+\e_2)/2.$ (~The equation is also
satisfied by $v=\infty$ which is the limit of one solution of
(\ref{ex}) as $\alpha\rightarrow 0$. However, such infinite solutions are 
trivial in the sense that they do not contribute to the eigenvalues
(\ref{eig}) for the conserved operators.~) For general values of $D$ with
$M=1,\,N=0$, (\ref{bae}) gives rise to a polynomial
equation of order $D$ for $\alpha\neq 0$, but this equation reduces to
order $(D-1)$ when $\alpha=0$. Whether 
the Bethe ansatz solutions give the complete spectrum for generic 
$\alpha$ is an
open problem
still to be solved, but the discussion above shows that it is 
possible since the breaking of the $so(5)$ symmetry to $so(3)$ is
accompanied by an increase in the number of solutions to the Bethe ansatz 
equations. For the purpose of counting the states, the results discussed in 
\cite{gk} may be appropriate.   

A final aspect for consideration is the possibility to compute form factors 
and correlation functions for this model. By rederiving the solution 
in the framework of the QISM, we hope to motivate further studies that
are necessary for this task, such as an analogue of Slavnov's formula 
for wavefunction scalar products, 
which is well known for $su(2)$ models \cite{slavnov}.

\begin{flushleft}
\bf{Acknowledgements}
\end{flushleft}
We thank the Australian Research Council for financial support. 
Jon Links thanks Departamento de 
F\'{\i}sica Te\'orica, Universidad de
Zaragoza, Spain, for their hospitality.

\end{document}